\documentclass[%
 reprint,
 amsmath,amssymb,
 aps,
floatfix,
]{revtex4-2}

\usepackage{graphicx}
\usepackage{dcolumn}
\usepackage{bm}

\usepackage[english,french]{babel}
\usepackage[utf8]{inputenc}
\usepackage[T1]{fontenc}
\usepackage{setspace}
\usepackage{hyperref}
\usepackage[french]{varioref}
\usepackage{mathtools}
\usepackage{empheq}
\usepackage{amsmath}
\usepackage{amsfonts}
\usepackage{amssymb}
\usepackage{amsthm}
\usepackage{cancel}
\usepackage{xcolor}
\usepackage{epstopdf}
\usepackage{setspace}
\usepackage{varwidth}
\usepackage{adjustbox}
\usepackage{array}
\usepackage{float}
\usepackage{tcolorbox}
\usepackage{array}
\usepackage{hhline}
\usepackage{makecell}
\usepackage{soul}

\hypersetup{
colorlinks,
citecolor=blue,
linkcolor=blue
}

\date{\today}

\begin{document}
\title{Multiscale study of high energy attosecond pulse interaction with matter and application to proton--Boron fusion}
\author{X. Ribeyre}
\affiliation{CEntre Laser Intenses et Applications, Univ. Bordeaux-CNRS-CEA, UMR 5107 Talence 33405, France}
\author{R. Capdessus}
\affiliation{CEntre Laser Intenses et Applications, Univ. Bordeaux-CNRS-CEA, UMR 5107 Talence 33405, France}
\author{J. Wheeler}
\affiliation{DER-IZEST, Ecole Polytechnique, 91128 Palaiseau Cedex, France}
\author{E. d'Humi\`eres}
\affiliation{CEntre Laser Intenses et Applications, Univ. Bordeaux-CNRS-CEA, UMR 5107 Talence 33405, France}
\author{G. Mourou}
\affiliation{DER-IZEST, Ecole Polytechnique, 91128 Palaiseau Cedex, France}
\email{}
\begin{abstract}

For several decades, the interest of the scientific community in aneutronic fusion reactions such as proton--Boron fusion has grown because of potential applications in different fields. Recently, many scientific teams in the world have worked experimentally on the possibility to trigger proton--Boron fusion using intense lasers demonstrating an important renewal of interest of this field. It is now possible to generate ultra--short high intensity laser pulses at high repetition rate. These pulses also have unique properties that can be leveraged to produce proton--Boron fusion reactions. In this article, we investigate the interaction of a high energy attosecond pulse with a solid proton--Boron target and the associated ion acceleration supported by numerical simulations. We demonstrate the efficiency of single--cycle attosecond pulses in comparison to multi--cycle attosecond pulses in ion acceleration and magnetic field generation. Using these results we also propose a path to proton--Boron fusion using high energy attosecond pulses.
\end{abstract}
\maketitle
\renewcommand{\labelitemi}{$\bullet$}
			    
\section{Introduction}
The application of lasers has revolutionized many fields in science and technology \cite{BretenakerBook}. But specifically short--pulse, intense lasers \cite{Mourou2006,Di-Piazza2012,Corde2013} are unique for efficiently delivering a coherent high peak power with a minimal amount of energy. 
This paper considers the application of coherent soft x-ray attosecond (10$^{-18}$ s) pulses of light to a nuclear process, namely as the driver for kinetic, athermal fusion within solid proton-Boron ($\rm p$--$^{11}\mathrm{B}$) targets.

In seeking higher peak intensity laser pulses, the fundamental Fourier limit for focusing the beam is defined by the wavelength ($\lambda_L$). The pulse intensity $\rm \left(I_L\right)$ is proportional to the pulse energy $\rm \left(\mathcal{E}_L\right)$ and inversely proportional to the focal volume$:$
$\rm I_L \propto \mathcal{E}_L/\lambda_L^{3}$. Here the laser pulse’s smallest volume, the so--called lambda--cubed regime \cite{Naumova2004}, is focused in space $\sim \lambda_L^{2}$ and time $\sim \lambda_L/\rm c$, where c is the speed of light.
Current laser amplification technology reaching the highest pulse energies at PetaWatt (PW$:$ $10^{15}$ W) laser facilities places the laser wavelength within the near--infrared (NIR) region of the spectrum ($\lambda_L \sim 1 ~\rm \mu m$; $T_{L}~\sim 3 ~\rm fs$) with femtosecond (10$^{-15}$ s) pulse durations ($\tau_L$) typically on the order of 25 to 150~fs. A transition to shorter wavelengths in the UV or X--ray region of the spectrum (UV--X)-- even at modest energies comparable to the current laser pulse energies of 10 to 100~J-- permits a dramatic increase in the peak pulse intensities attainable.
Progress in short--wavelength short pulse sources have been made, i.e. with renewed interest in X--Ray free electron lasers (XFEL) for the production of attosecond pulses~\cite{Shim2020}. However, production from sub--cycle laser--driven processes such as high--order harmonic generation sources in gas or solid \cite{Chatziathanasiou2017} show the greatest promise for achieving such ultra--short durations with sufficient energy.

This arises from the exploration of laser--plasma interactions with few-cycle pulses in relativistic intensity (several $10^{19}~\rm W/cm^2$) NIR laser facilities. It is now possible to generate these relativistic intensity laser pulses at high repetition rate (kHz) with only a mJ of energy by focusing to the lambda--cubed regime~\cite{Ouille}. The control over high intensity, few--cycle NIR pulses has the potential to achieve considerable breakthroughs in high efficiency laser particle acceleration~\cite{Zhou, Wu}, high efficiency and high energy radiation sources, ultra--high amplitude magnetic field generation and ultra-high pressures. There are many PW--scale laser facilities being built around the world~\cite{Danson2015} that are operating at 1~shot/min rates and are now targeting repetition rates of 1~Hz. This represents a dramatic increase of earlier technological capabilities that required many minutes and up to several hours of recovery time between subsequent shots. This progress in operational repetition rate can be expected to continue with the introduction of new laser architectures and technologies such as the coherent amplification network (CAN) laser ~\cite{Mourou2013, Soulard2015, Wheeler2016, Fsaifes2020} and thin disk amplifiers~\cite{Fattahi2014, Reagan2018}. 

Despite the growing number of PW laser facilities, to reach few--cycle NIR pulse durations, they will require high--energy, post--compression techniques.
The thin film compressor~\cite{Mourou2014, Khazanov2019} works for pulses with energies of 10 to 100~J with minimal losses in order to most efficiently drive laser--plasma interactions. These NIR post--compression techniques have been studied at the TW--scale~\cite{Mironov2017,Farinella2019,Ginzburg2020} and work is now implementing the method on PW--scale laser systems~\cite{Ginzburg2021}.  
The production of few--cycle NIR pulses will offer the tools for efficient relativistic compression to UV--X attosecond pulses through laser--plasma interactions such as the relativistic plasma mirror (RPM) \cite{Bulanov1994,Lichters1996,Quere2008,Vincenti2019,Chopineau2021} and will make it possible to explore laser--plasma interactions with these attosecond UV--X pulses in solid--state target media under novel conditions~\cite{Tajima2014, Wheeler2019}.
The corresponding photon energy range for these attosecond pulses is from 10~eV to 1~keV which can propagate even inside solid density targets. 
Another unique feature that will contribute to this ultra--high intensity regime 
will be the additional contribution due to the self--focusing effects arising from the non--linear interaction of these attosecond pulses with matter. Their study enables the exploration of new regimes of laser--matter interaction with a plethora of potential applications in fundamental physics and extreme laboratory astrophysics with links arising to high field quantum electrodynamics, nuclear physics and general relativity. These UV--X pulses also have unique properties that can be leveraged to produce proton--Boron (pB) nuclear fusion reactions.

\begin{figure}
    \includegraphics[scale=0.4,angle=0,clip]{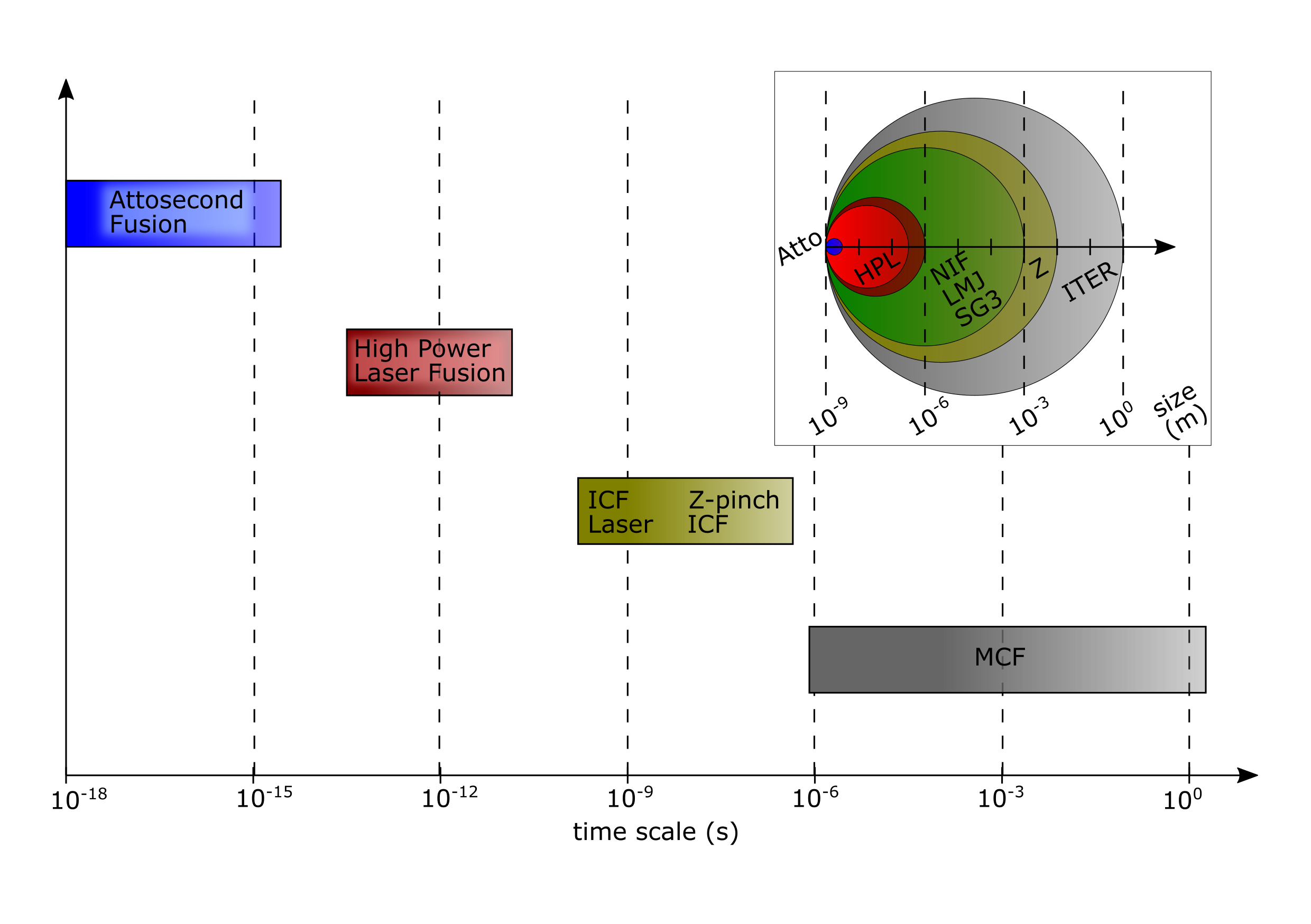}
    \caption{Landscape of nuclear fusion with the corresponding fusion schemes and facilities on different time and space scales i.e. from attosecond fusion to magnetic confinement fusion.}\label{Fusion_scope}
\end{figure}

Different ways have been investigated to achieve controlled fusion reactions. All these schemes explore different physical regimes (see Fig.~\ref{Fusion_scope}), covering vastly different scales in space and time$:$ i) Magnetic Confinement Fusion (MCF) \cite{MCF} requires timescales of multiple seconds at the size scale of meters, ii) Inertial Confinement Fusion (ICF)  with lasers or Z--pinches \cite{ICF,ICF_Pinch} fall between the scales of 1--100~ns and 1--10~mm, iii) high--power laser fusion \cite{Hora} are in the picosecond and micrometer scales, while  iv) attosecond fusion, which mainly concerns this paper, falls within the scale of attoseconds and nanometers.

We would like to emphasize that there is a fundamental difference between the fusion schemes at the long timescale, which includes everything greater than a picosecond ($> 10^{-12} s$), and the short timescale, which includes attosecond and femtosecond. In the case of long timescale fusion, the plasma is close to the Maxwellian equilibrium, and is referred to as thermonuclear fusion. In the case of the short timescale, the particle energy distributions are out of equilibrium and no longer in a Maxwellian distribution. Such a scheme is then considered athermal fusion. 

\begin{figure}
    \includegraphics[scale=0.26,angle=0,clip]{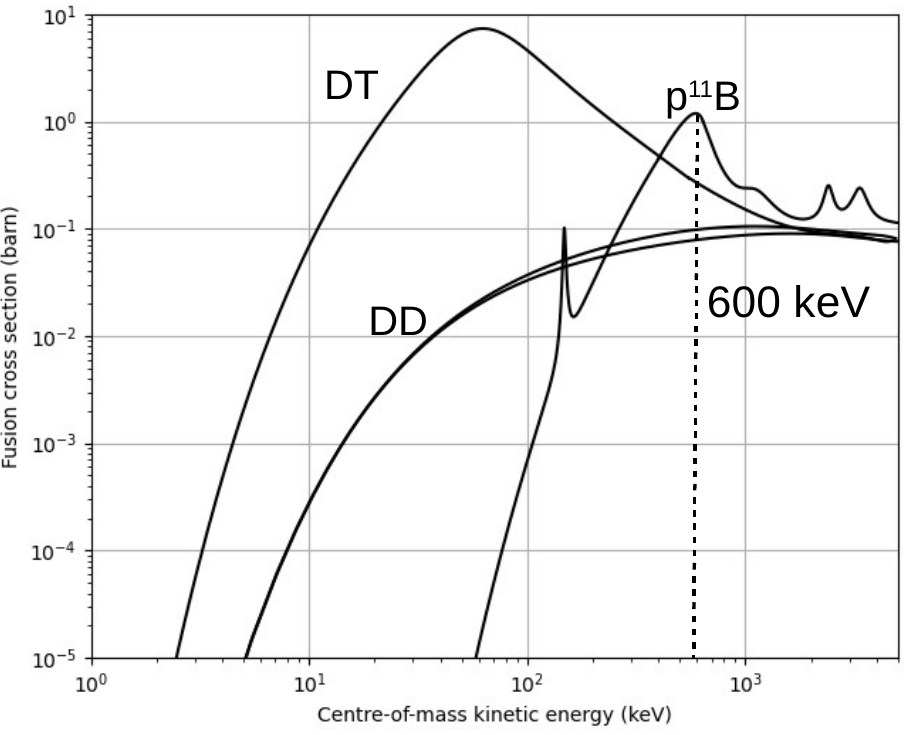}
    \caption{Fusion cross sections of Deuterium--Tritium (DT), Deuterium--Deuterium (DD) and proton--Boron (p--$^{11}$B) as a function of the center of mass kinetic energy. The dotted line indicates the energy of the higher cross section resonance ($\sim$~600 keV). For DT and DD fusion cross sections, see Ref.~\cite{Bosch_Hale} and for pB fusion, see Ref.~\cite{Nevins_Swain}.\label{cross_section}}
\end{figure}

For several decades, the interest of the scientific community in aneutronic fusion reactions such as pB fusion has grown because of potential applications in various fields${:}$ alpha particle sources, cancer treatment, radio-isotope production, fusion propulsion or clean neutronless fusion reactors. Despite the low cross-section, 1~barn, near the main resonance ($ \rm E_{cm} \simeq 600~\rm keV$) for pB compared to 6~barns at $\rm E_{cm} \simeq 60~\rm keV$ for deuterium--tritium (DT), where $\rm E_{cm}$ is the center of mass frame energy, the pB fusion cross section does not decrease significantly in the MeV range (see Fig. \ref{cross_section}). Proton--Boron fusion $\rm B(p,\alpha)2\alpha$ is neutronless while deuterium--tritium fusion $\rm D(T,\alpha)n$ produces high energy neutrons (14~MeV) which are difficult to stop, making it difficult to avoid activation of the surrounding material. 
However, if we compare pB fusion and DT fusion in terms of burning mass, to produce 10 J of fusion energy, assuming the pB density at $1~\rm g/cm^3$ (solid density) and for DT 0.25~$\rm g/cm^3$ (solid cryogenic DT) the burning volumes are quite comparable in the two cases i.e. 140~$\rm \mu m^3$. Assuming a cylindrical volume with a typical length of 20~$\rm \mu m$, the corresponding radius of the burning volume is 1.5~$\rm \mu m$. We choose 10~J of fusion energy production because this is the energy envisaged for high energy attosecond pulses. Such pulses will be able to penetrate solid density pB targets over $ \sim$~20~$ \rm \mu m$ and will be focused on $\sim$ 100~$\rm nm$. The interaction volume can therefore be represented as a cylinder of 0.05 $\rm \mu m$ radius and 20~$\rm \mu m$ long. The interaction volume of such pulses is $\sim$ 30~times less in radius (if we still consider a cylindrical volume) than the burning volume. 
To study an application to energy production, all the physical processes initiated by the attosecond pulse that might contribute to burn such a volume must be considered$:$ interaction and propagation of the pulse, acceleration of protons and Boron ions, and propagation of these ions.

Recently, many scientific teams in the world have worked experimentally on the possibility to trigger pB fusion using intense lasers. In 2005 Belyaev \textit{et al.} \cite{Belyaev} used a picosecond laser pulse with an intensity of about $10^{18}~\rm W/cm^2$ interacting with a Boron--rich target to produce $10^3~\rm \alpha/sr/shot$ and later Kimura \textit{et al.} estimated that this number was underestimated and in fact reached $10^5~\rm \alpha/sr/shot$ in 2009 \cite{Kimura}. In 2013, Labaune \textit{et al.} \cite{Labaune} demonstrated a maximum alpha-particle yield of 9$\times10^6~\rm \alpha/sr/shot$. Picciotto in 2014 \cite{Picciotto} achieved $10^9~\rm \alpha/sr/shot$ with a similar set-up but with a nanosecond laser. And very recently Giuffrida~\cite{Giuffrida} obtained above $10^{10}~\rm \alpha/sr/shot$ and demonstrated the non-thermonuclear nature of the $\alpha$ production. All these works demonstrated the important renewal of interest of this field, and have led to new studies on pB nuclear fusion~\cite{Hora}. 

In this article, through advanced particle--in--cell (PIC) simulations and modelling, we demonstrate the efficiency of a single--cycle pulse for radial ion acceleration and generation of mega Tesla--level magnetic fields via the interaction of an ultra-high intense attosecond pulse with a solid target. In doing so, we can propose a new path with optimal conditions for energy production by nuclear fusion \cite{Patent}.

\begin{figure}
    \includegraphics[scale=0.5,angle=0,clip]{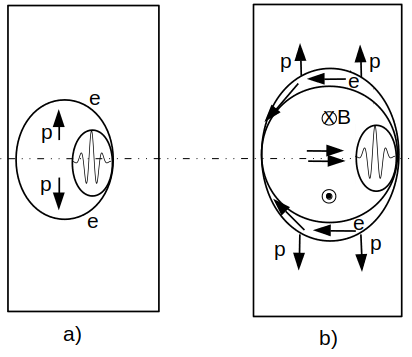}
    \caption{Sketch of the scenario of the pulse laser--matter interaction. (a) The action of the laser pushes radially the electrons (denoted $e$) which creates a positively charged channel. The protons ($p$) are then accelerated by the induced electric field. (b) The high density electron current generates a strong toroidal--shape magnetic field ($\textbf{B}$). The electron charge accumulation on the channel border allow to re--accelerates the protons.}\label{scenario}
\end{figure}
To explain the main steps of the interaction of an ultra-intense attosecond pulse with a solid target, the involved physical processes are showed in Fig.~\ref{scenario}. On Fig. \ref{scenario}a the pulse penetrates inside the solid target because at 10~nm range (wavelength) the solid target is under critical. The transverse ponderomotive force expels the electrons from the channel and generates a strong electric field. This electric field accelerates ions (protons) transversely. Figure \ref{scenario}b shows how the accumulated electrons on the channel boundary can re--accelerate the ions (see the two ellipse curves). Moreover, the electron current generates a strong toroidal magnetic field. These figures help to have an overview of the physical processes that we plan to study in this paper. 

The remainder of the article is organized as follows.
In Section \ref{Sec2}, the parameters of the PIC simulations supporting our study are detailed. Section \ref{Sec3} describes the key events of the interaction of an attosecond pulse with a solid density target from the numerical simulation results supported by analytical estimates. In particular, we show the efficiency of a single--cycle pulse (compared with a 10--cycle pulse) for ion acceleration and the generation of mega Tesla--level magnetic fields. In Section \ref{Sec4} we propose a path to proton--Boron fusion using high energy attosecond pulses. Section  \ref{Sec5} is devoted to our conclusions and some perspectives.

\section{Simulation setup} \label{Sec2}
We have performed 2D3V (i.e. two-dimensional in configuration space, three-dimensional in momentum space) Particle-In-Cell (PIC) numerical simulations using the EPOCH code \cite{Arber} describing the interaction of a linearly polarized pulse with a proton--Boron (pB) plasma, to investigate this new interaction regime. The laser field profile is Gaussian in time and space with a beam waist $\left(w_0 \right)$ of 60 nm. The maximum laser intensity is 2.71$\times 10^{26} ~\rm W/cm^{2}$, corresponding to a normalized potential vector $a_0 \equiv eE_L/m_ec\omega_L$=140 where $E_L$, $\omega_L$ and $m_e$ are the laser electric field strength, angular frequency and the electron mass, respectively. The initial laser field as a function of $a_0$ can be written :
\begin{align}\label{profile}
{\bf a}_L\left(y,t\right) = a_0 \exp \left[- \frac{\left( t-t_{\text{rise}}\right)^2}{T_L^2} \right] \exp\left[- \frac{y^2}{w_0^2} \right] \sin\phi {\bf\,\, e}_y
\end{align}
where $\phi= \omega_L t-k x$ and $t_{\text{rise}} \simeq 2~ \tau_L$ are the phase and the rise time of the temporal profile, respectively. Here, $T_L= \sqrt{2/\pi}~\tau_L $ where $\tau_L$ is the laser pulse duration.

The target is composed of Boron ions and protons and is fully ionized (solid density${:}$ 1 g/cc). We use an electron density profile defined as $n_e~=~n_{eo}/2\left\{\tanh\left[2\left(x-x_d \right)/x_d \right]+1\right\}$ where $x_d= 10~\rm \lambda_L$ and $n_{e0}= 3\times10^{23}~\rm cm^{-3}$ is the (maximum) initial electron density. Such a value corresponds to $n_{e0}\simeq 0.03~n_c$, where $n_c= \epsilon_0m_e\omega_L^2/e^2$ is the critical density at the laser wavelength $\lambda_L=$~10~nm, $\epsilon_0$ is the vacuum permittivity and $e$ is the elementary charge. While such a dense plasma would be opaque to mildly relativistic laser pulses at laser wavelengths of $\sim$1~$\rm \mu m$, it turns transparent at the laser wavelength considered here. The laser pulse starts irradiating the plasma at $t=0$. 

The simulation box is defined by a spatial grid of dimensions $900\times 1500$~nm$^2$ using $2250\times3750$ mesh cells, and the target is represented by 3.51$\times 10^8$ macroparticles.


To study the influence of a single--cycle laser pulse  (i.e. $\mathcal{E}_L$~=~1 J and $\tau_L$~=~$1~T_L$ ) on the collective dynamics of the plasma, comparisons will be made with a 10--cycle laser pulse (i.e. $\mathcal{E}_L$= 10 J and $\tau_L=10~T_L$), where $T_L~=~\lambda_L/c=~33$ as is the laser period.


\section{Simulation results and analytical estimates}\label{Sec3}
\subsection{Overall interaction scenario}\label{Sec3a}
\begin{figure}
    \includegraphics[scale=0.47,angle=0,clip]{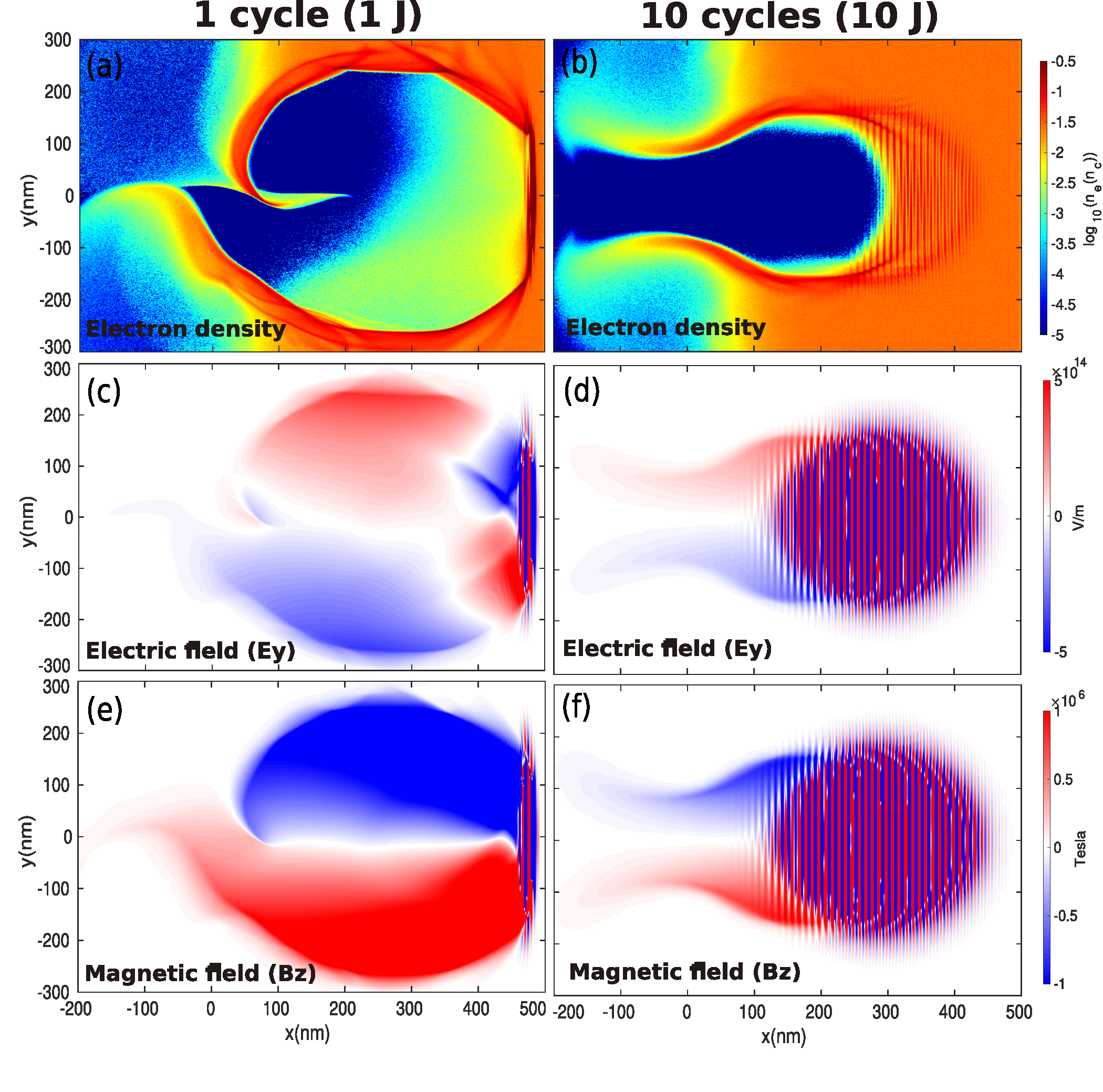}
    \caption{Maps of the electron density (a) and (b), electric field ($E_y$) (c) and (d)  and magnetic field ($B_z$) (e) and (f) ~at $t=50~\rm T_L \simeq 1.65$ fs.}\label{densities_fields_50}
\end{figure}

At the beginning of the interaction, the electrons are efficiently expelled from the center of the propagation axis (i.e. the $x$--axis), forming a channel quasi-devoid of electrons as shown in Figs.
\ref{densities_fields_50}(a) and \ref{densities_fields_50}(b) for a single--cycle laser pulse and a 10--cycle laser pulse, respectively. Clearly, the duration of the pulse $\left( \tau_L\right)$ has a significant influence on the generation of the self-consistent fields as shown in Figs. \ref{densities_fields_50}(c)-(f). With a single--cycle pulse, the radius of the channel is increased by $\approx$ 50\%, even though the laser energy is 10 times less. How is such an effect possible$?$ This can be explained through two mechanisms that eject the electrons from the highest field regions$:$ (1) Via the longitudinal and transverse gradients of the laser electromagnetic wave, i.e. by the action of the so--called ponderomotive force$;$ (2) Via a \textit{kick} directly induced by the laser electric field itself. For long laser pulses (i.e. $\tau_L \gg T_L$),  mechanism (1) dominates and is responsible for the particle ejection. 
However, for $\tau_L \sim T_L$, the electron ejection is primarily caused by mechanism (2).  

\begin{figure}
    \includegraphics[scale=0.47,angle=0,clip]{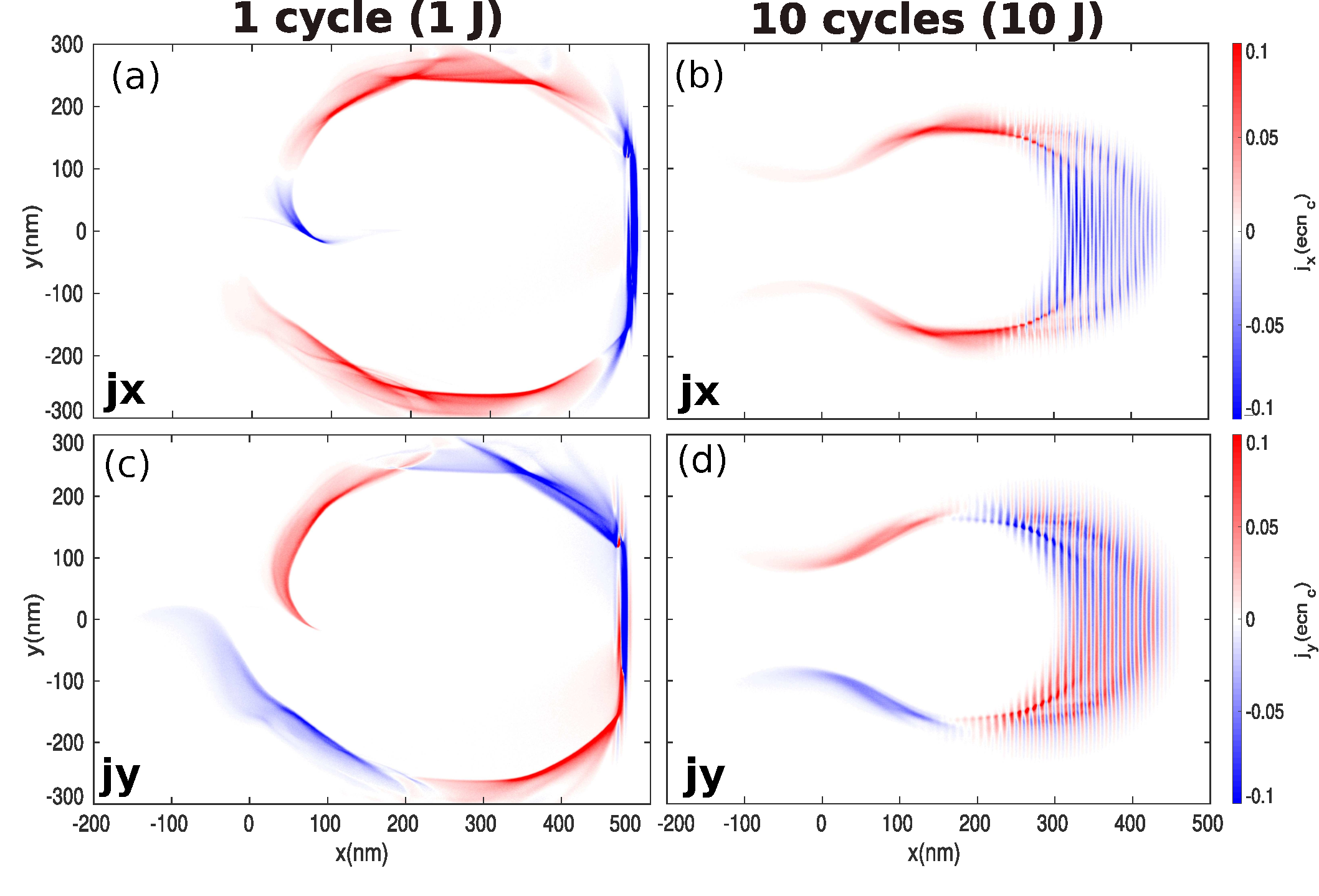}
    \caption{(a) and (b) longitudinal and  (c) and (d) transverse electron currents at $t=50~T_L$.}\label{currents}
\end{figure}

It has been shown that the excursion of the electrons from the regions of highest fields can be expressed as~\cite{Hartemann_PRE}$:$
\begin{subequations}
\begin{align}
 y_e\left( \phi \right) = \gamma_{e0}\left(1+\beta_{e0} \right)\left[ A\left( \phi\right) + B \left( \phi\right) +q\phi \right] \,
\end{align}
\end{subequations}
where $q= \kappa^2 p$ with $p= a_0 \kappa/\left[2\left(1-\kappa^2 \right)\right]$, $A$ and $B$ are defined by Eqs. (49) and (50) in Ref. \cite{Hartemann_PRE}.  
The parameter $\kappa$ is linked to the number of cycles $\left( N\right)$ via the following expression$:$
\begin{equation}
\kappa= \frac{\pi}{\omega \tau_L} = \frac{1}{2N}.    
\end{equation}
In the vicinity of the focus (i.e. $x \ll x_0= \pi w_0^2n/\lambda_L $, where $x_0$ is the Rayleigh length with $w_0$ and $n$ the laser beam waist and the refractive index, respectively), the quiver amplitude can exceed the radius of the laser focal spot. Therefore, as the electrons quiver through the spatial gradient of the laser fields, the restoring force decays exponentially and the electrons are scattered away from the focus with high energies. This induces the generation of suprathermal (i.e. non-Maxwellian) electrons. Taking the finite beam waist into account, the excursion radius $\left( r_y\right)$ can be estimated as$:$
\begin{align}
    r_y \sim q \lambda_L
     = \left\{
    \begin{array}{ll}
        a_0 \lambda_L/6  & \mbox{N=1}, \\
     \mathcal{O}\left( \kappa^2\right)  & \mbox{N=10}.
    \end{array}\label{kappa1}
\right.
\end{align} 
Equation \eqref{kappa1} means that a single--cycle pulse drastically enhances the effect of mechanism (2) which can be seen as a nonlinear ponderomotive scattering effect. In the case of a single--cycle pulse, the electrons can reach  the maximum electric field within half a laser period and are \textit{immediately} scattered away radially.
Figure~\ref{currents} shows the longitudinal and transverse electron currents. For a single--cycle pulse, the longitudinal \textit{ponderomotive force}~\cite{comment} is strongly enhanced which tends to increase the forward-directed electron current (compared with a 10--cycle laser pulse), up to $\approx -e n_c c$ as shown in Fig. \ref{currents}(a) (note that the color scale is saturated for this diagnosis). Due to the charge conservation, return currents are generated at the edge of the channel. 
Figures \ref{currents}(c) and \ref{currents}(d) show the transverse electron currents at $t=50~\rm T_L$. We can notice the effect of mechanism (2) which tends to enhance the amplitude of the transverse electron current as the electrons are efficiently scattered away by the  electric field of the laser itself with the use of a single--cycle pulse. 

We can therefore deduce the average radius of the channel $r_{\text{ch.}}$: 
\begin{align}
r_{\text{ch.}}= \max\left\{r_y; r_0\simeq w_0\sqrt{\ln \left[\frac{1}{\pi^2}\frac{a_0}{\sqrt{2}}\frac{n_c}{n_{e0}}\frac{\lambda_L^2}{w_0^2} \right]} \right\},
\end{align}\label{radius}
where $r_0$ can be obtained from a balance between the electrostatic force and the ponderomotive force. For $N~=~1$, $r_{ch.}=r_y\simeq 240$~nm (see Eq.~\ref{kappa1}) and for $N~=~10$, $w_0\simeq$~6 $\lambda_L$, $n_{e0}=0.03~n_c$, then $r_{\text{ch.}}= r_0\simeq 90$ nm. Those estimates are in good agreement with simulation results  -- albeit the channel radius $\left( r_0\right)$ is a bit underestimated in the case of a  pulse of 10--cycles-- as shown  in Figs.~\ref{densities_fields_50}(a) and \ref{densities_fields_50}(b). 

Thus, the maximum amplitude of the induced azimuthal magnetic field $\left(B_z \right)$ can be estimated from Ampere's law $\boldsymbol{\nabla}\times {\bf B} =\mu_0 {\bf j}$ as$:$
\begin{align}\label{Bz}
    \max\left[B_z \right]&\sim 2\pi\frac{n_{e0}}{n_c}\frac{r_{\text{ch.}}}{\lambda_L}\frac{\omega_L m_e}{e} \\
    & \simeq \left\{
    \begin{array}{ll}
        5 ~\text{MT} & \mbox{N=1} \nonumber\\
        3.5 ~\text{MT} & \mbox{N=10}.\nonumber 
    \end{array}
\right.
\end{align}
The self--consistent magnetic field $\left( B_z \right)$ has a strong influence on the electron acceleration (e.g. see Ref.~\cite{Pukhov_1998} where the authors refer to it as the \textit{B-loop} mechanism). This B-loop mechanism can be identified via the topology of electron currents, shown in Fig. \ref{currents}. In our case, it is expected that the amplitude of the toroidal magnetic field $(B_z)$ can reach several mega Tesla (MT).
Moreover, as $B_z$ has a nearly linear dependence on $(r_\text{ch})$ and reaches its maximum at the inward edge of the channel ($y= r_{\text{ch.}}$), its amplitude is strongly enhanced with a single-cycle laser pulse (see Eq. \eqref{Bz}), which is consistent with numerical simulation results (see Figs. \ref{densities_fields_50}(e) and \ref{densities_fields_50}(f)).
\subsection{Radial ion acceleration}

It is now interesting to turn our attention to the radial ion acceleration. Here, the \textit{radial} ions are those where the momentum components satisfy the condition $\vert p_{iy}\vert \geq \vert p_{ix} \vert$.
\subsubsection{Regime of ponderomotive acceleration}
In the case of a laser pulse duration $\tau_L$ longer than the plasma electron period ($\sim 1/\omega_{pe}$~=~30~as), the electrons respond adiabatically. In our case for $\tau_L~=~10~T_L =~330$~as, this condition is fulfilled. The transverse ponderomotive force, $ {\bf F}_{p e} = - \partial_y \gamma_e m_ec^2{\bf e}_y$ induces an electric field, ${\bf E} = - e n_e/\left( 2\epsilon_0\right) {\bf y}$, due to the charge separation. At mechanical equilibrium, i.e. in the so--called regime of ponderomotive acceleration \cite{Macchi_2009}, we have $ {\bf F}_{pe}=e  {\bf E}$. 
Consequently, the equation of motion for radially accelerated ions can be recast as \cite{Sarkisov}${:}$  
\begin{equation}\label{equ:motion}
\frac{d  v_{ir}}{d t}=-\frac{Z m_e c^2}{A m_p} \partial_y \sqrt{1+\langle {\bf a}_L^2\left(y,t\right)\rangle},
\end{equation}
where $A$ and $m_p$ are the ion mass number and the proton mass, respectively. In Eq. \eqref{equ:motion}, the brackets denote an average over the laser period such that$:$
\begin{align}
\langle {\bf a}_L^2 \left(y,t \right) \rangle &=\langle \alpha \rangle a_0^2 \exp\left[-\frac{2\left( t-t_{\text{rise}}\right)^2}{T_L^2}\right]\exp\left[-\frac{2y^2}{w_0^2} \right] ,\\
&\langle \alpha \rangle \equiv  \frac{1}{T_L}\intop_{t_{\text{rise}}}^{t_{\text{rise}}+T_L}  \sin^2\phi\exp\left[-\frac{2\left(t-t_{\text{rise}}\right)^2}{T_L^2} \right] dt. \label{a_carre}
\end{align}
In specific limits we have$:$  
 \begin{align}
 \langle \alpha \rangle  \simeq \bigg\{
    \begin{array}{ll}
         1/2  & \mbox{$\tau_L \gg T_L$}  \\
         0.37 & \mbox{$\tau_L = T_L$,}     
    \end{array}
 \end{align}
 where the first limit is related to the slowly varying envelope approximation (i.e. $\partial_t {\bf a}_L \ll \omega_L {\bf a}_L$) used in most papers involving laser ponderomotive aspects.
Since the target thickness is much smaller than the depletion length of the laser electric field, it is reasonable to assume that the spatio--temporal profile of the laser electric field is weakly disturbed during its propagation within the plasma. By injecting Eq. \eqref{profile} into Eq. \eqref{equ:motion}, using the slowly varying envelope approximation and $a_0 \gg 1$, it can be shown that the maximum velocity for the radially-accelerated ions can be written$:$
\begin{equation}\label{equ:Vi}
\max\left[ v_{ir}  \right]\simeq K\frac{Zm_e c^2}{Am_p} \frac{\tau_L}{w_0}a_0
\end{equation}
where $K= \exp\left( -1/2\right)\simeq 0.6$.
Then from Eq. \eqref{equ:Vi}, and assuming non--relativistic ions, the expression for the maximum ion kinetic energy can be recast as${:}$
\begin{equation}\label{equ:Emax}
 \max \left[ E_{ir}\right] \simeq \frac{1}{2}\frac{Z^2}{A} \frac{m_e}{m_p}\left[a_0\frac{\tau_Lc}{w_0}\right]^2m_ec^2
\end{equation}

It is worth noting that we find about a factor two times more for the kinetic ion energy obtained by Sarkisov \textit{et al.} \cite{Sarkisov}, because their solution fits for $a_0$ close to unity. We recall that expression \eqref{equ:Emax} holds for relatively long pulses, i.e. in the weakly time varying pulse envelope but can be applied to our study in the case of a 10--cycle pulse (where the factor $\tau_L/w_0$ is equal to 1/c).
Therefore, the maximum proton energy is $\max\left[E_{pr}\right] \simeq 2.9~\rm MeV$ and the maximum energy for the Boron ions is $\max\left[E_{Br}\right]\simeq 6.8 ~ \rm MeV$ (see Eq. \eqref{equ:Emax}). Such estimates are in good agreement with the simulation results as shown in Fig. \ref{radial_plots}(a) until $t \approx 70~T_L$. At longer times, other processes of acceleration become important and tend to enhance the maximum average ion kinetic energy but are not taken into account by the Sarkisov model. This is due to the fact that even with a 10--cycle pulse the radius of the channel slightly increases over time, which tends to break the key hypothesis of the Sarkisov model i.e. a mechanical equilibrium between the transverse ponderomotive force and the radial electrostatic force. 

\subsubsection{Beyond the regime of ponderomotive acceleration}
\begin{figure}
    \includegraphics[scale=0.42,angle=0,clip]{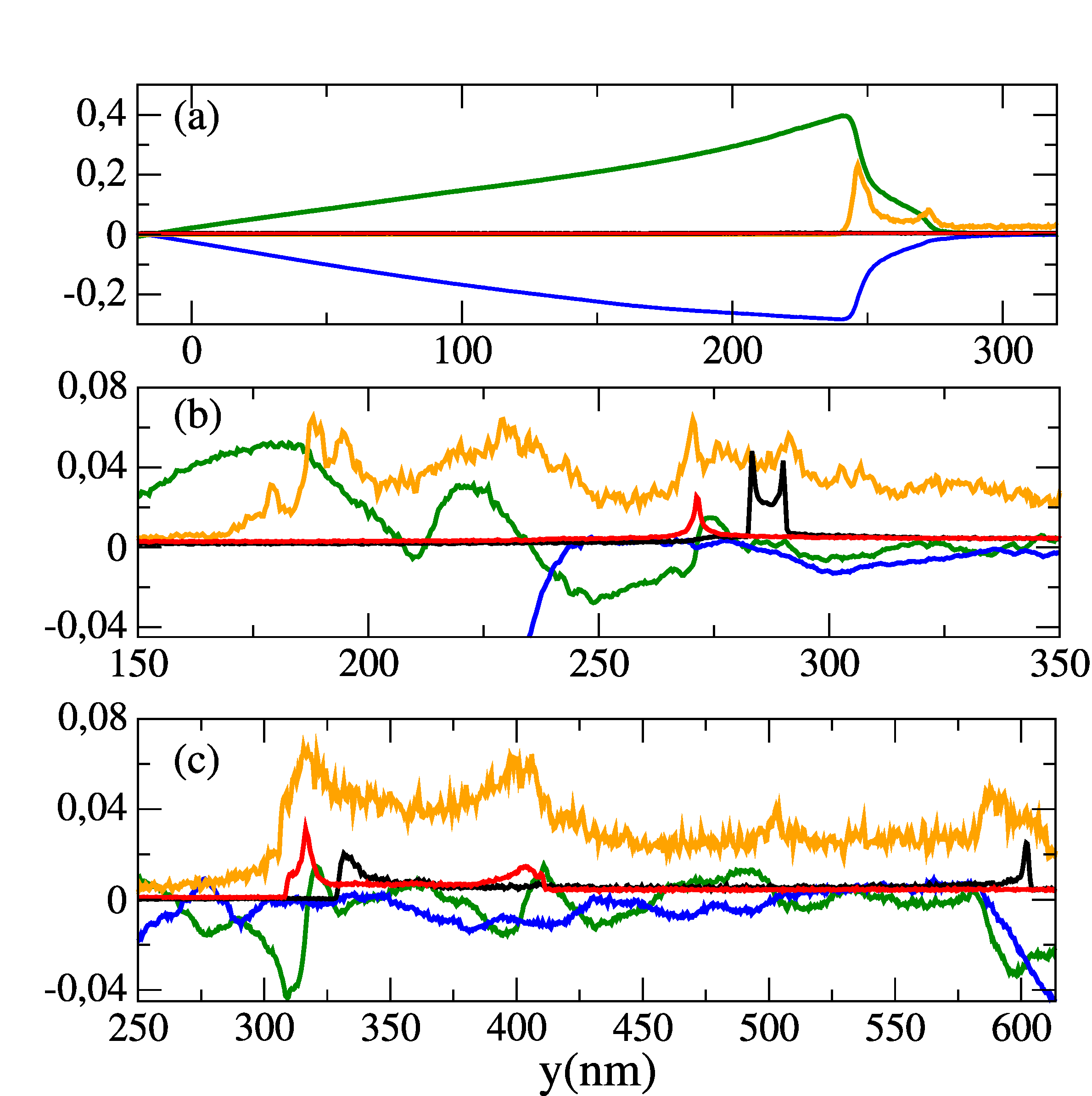}
    \caption{Lineouts of the electron ($n_e(n_c)$, orange) and proton ($n_p(n_c)$, black) and Boron ($n_b(n_c)$, red) number densities, and of the longitudinal electric field ($E_y(1\times 10^{15}~\rm \text{V/m})$, green) and magnetic field ($B_z$(5$\times 10^6~\rm \text{T})$, blue) along $x=200$ nm, at (a) t= 50~$\rm T_L$, (b) t= $90~\rm T_L$ and (c) $t= 250~\rm T_L$.}\label{1D_cuts}
\end{figure}

In the case of a single--cycle pulse, estimate \eqref{equ:Emax} is no longer valid as the aforementioned mechanical equilibrium no longer exists due to a complete electron depletion in the channel by the electric field itself. Instead, the radially accelerated ions satisfy the following equation of motion$:$
\begin{align}\label{ion_eq}
    \frac{d^2y_i}{dt^2} =& \Omega^2 \{y_i \left[\mathcal{H}\left(y_i \right)- \mathcal{H}\left( y_i-r_{\text{ch.}}\right) \right]  \\
&+r_{\text{ch.}}\frac{l-y_i}{l-r_{\text{ch.}}}\left[\mathcal{H}\left(y_i -r_{\text{ch.}}\right)-\mathcal{H}\left( y_i-l\right)\right]  \}\nonumber, 
\end{align}
where $\mathcal{H}$ is the Heaviside function, $\Omega=\sqrt{e^2n_{e0}/2\epsilon_0 m_i }$ and $l= r_{\text{ch.}}+\delta r$. Here $\delta r$ is the thickness of the electron compression layer (see Fig. \ref{1D_cuts}(a)) such that $\delta_r~\sim~c/\omega_{pe}~\approx \lambda_L~\sqrt{\gamma_e}/\left(2\pi \sqrt{n_{e}}\right)~\simeq~6.5 ~\lambda_L$, in good agreement with simulation results.
The form of the electric field (see Eq. \eqref{ion_eq}) can be identified in Fig. \ref{1D_cuts}(a). It is worth stressing that Eq. \eqref{ion_eq} is only relevant for ions radially accelerated toward the edge of the channel, i.e. satisfying $y \geq r_{\text{crit.}}$ where $r_{\text{crit.}}$ is the critical radius. For $y < r_{\text{crit.}}$, the ions are pinched and forward accelerated \cite{Popov, Sahai}. We will ignore this ion population in the remainder of the manuscript, as it is not within the scope of the paper.

\begin{figure}
    \includegraphics[scale=0.45,angle=0,clip]{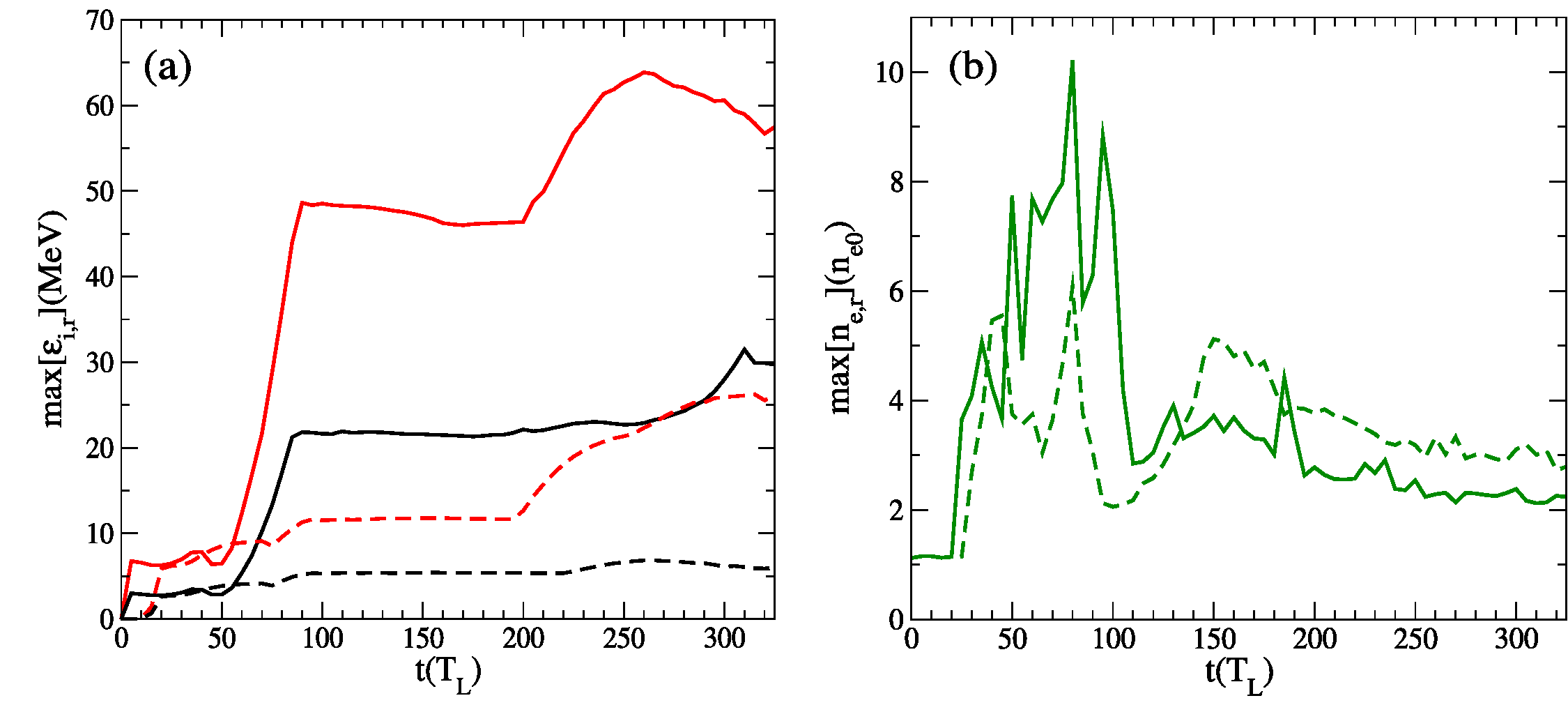}
    \caption{(a) Maximum radial ion energies and (b) maximum electron densities, as a function of time. In continuous (dashed) lines, with a single (10) cycle(s) pulse. In black (red), protons (Boron ions) and in green, electrons.}\label{radial_plots}
\end{figure}

\begin{figure}
    \includegraphics[scale=0.7,angle=0,clip]{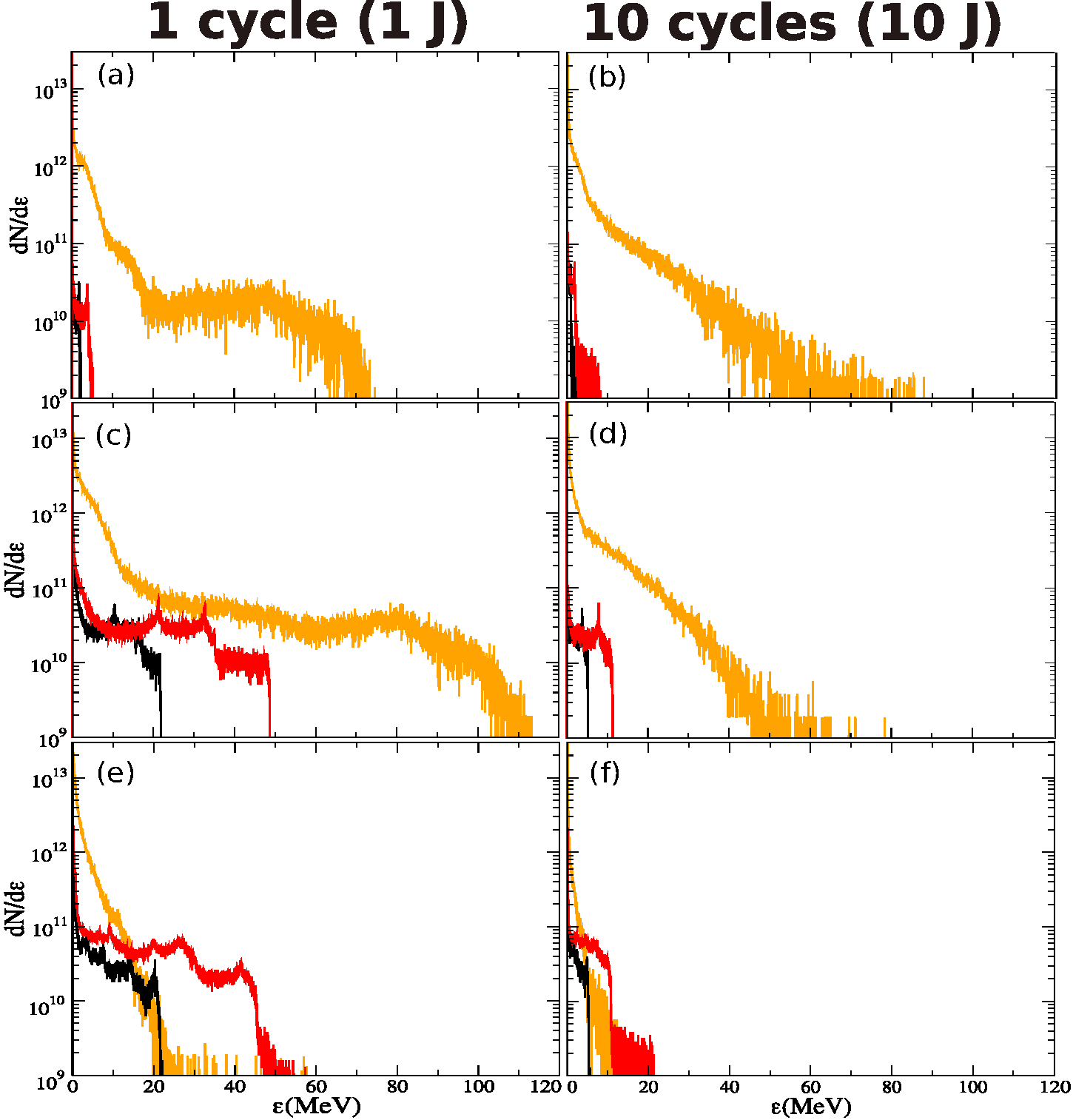}
    \caption{Radial energy spectra of electrons (orange), protons (black) and Boron ions (red) at (a)--(b) $t=50~\rm T_L$, (c)--(d) $t= 90~\rm T_L$ and (e)--(f) $t=250~\rm T_L$. Left column, with a single--cycle pulse. Right column, with a 10--cycles pulse.}\label{radial_energy_spectra}
\end{figure}

Figure \ref{radial_plots}(a) shows the evolution of the maximum radial ion kinetic energy as a function of the time.  
Up to $t \approx 50 ~T_L$, the ions are accelerated by the radial electrostatic force (modelled via the r.h.s of Eq. \eqref{equ:motion} in the case of a 10--cycle pulse and via the first r.h.s term Eq. \eqref{ion_eq} in the case of a single--cycle pulse). During this acceleration stage, most of the accelerated ions reach the same point which leads to the generation of a density spike of protons  -- and Boron ions later -- at $y_i \approx r_{\text{ch.}}$. For $50~T_L \lesssim t \lesssim 90~T_L$, the ions are further accelerated by the charge separation field of the electron compression layer (see second term of r.h.s of Eq. \eqref{ion_eq}). We note that the ions gain most of their kinetic energy in the electron compression layer. Moreover, contrary to the previous stage of acceleration (i.e. $t\leq 50~T_L$), the effect of the pulse duration is much more significant. For instance, the maximum proton energy is increased by a factor 10 (from $t\approx50~T_L$ to $t\approx90~T_L$)
whereas it is increased by at most 2 in the case of a 10--cycle pulse, as shown in Fig. \ref{radial_plots}(a). 
Figure \ref{radial_plots}(b) shows the evolution of the maximum electron densities as a function of time. We note that the maximum density during the second stage of acceleration (i.e. $50~T_L\leq t \leq 90~T_L$), which corresponds to the density of electron compression layer is larger in the case of a single--cycle pulse. This is consistent with the fact that the electrons are more efficiently  expelled radially, which therefore increases the channel radius (see section \ref{Sec3a}). This reinforces the robustness of the electron compression layer. Therefore, the charge separation field within the electron compression layer is stronger -- for a single--cycle pulse compared with a 10--cycle pulse -- as is also true for the maximum ion kinetic energy. Moreover, non--Maxwellian electrons are generated, unlike in the case of the 10--cycle pulse. This electron population can especially be identified by a quasi--plateau in Fig. \ref{radial_energy_spectra}(c). 

\begin{figure}
    \includegraphics[scale=0.47,angle=0,clip]{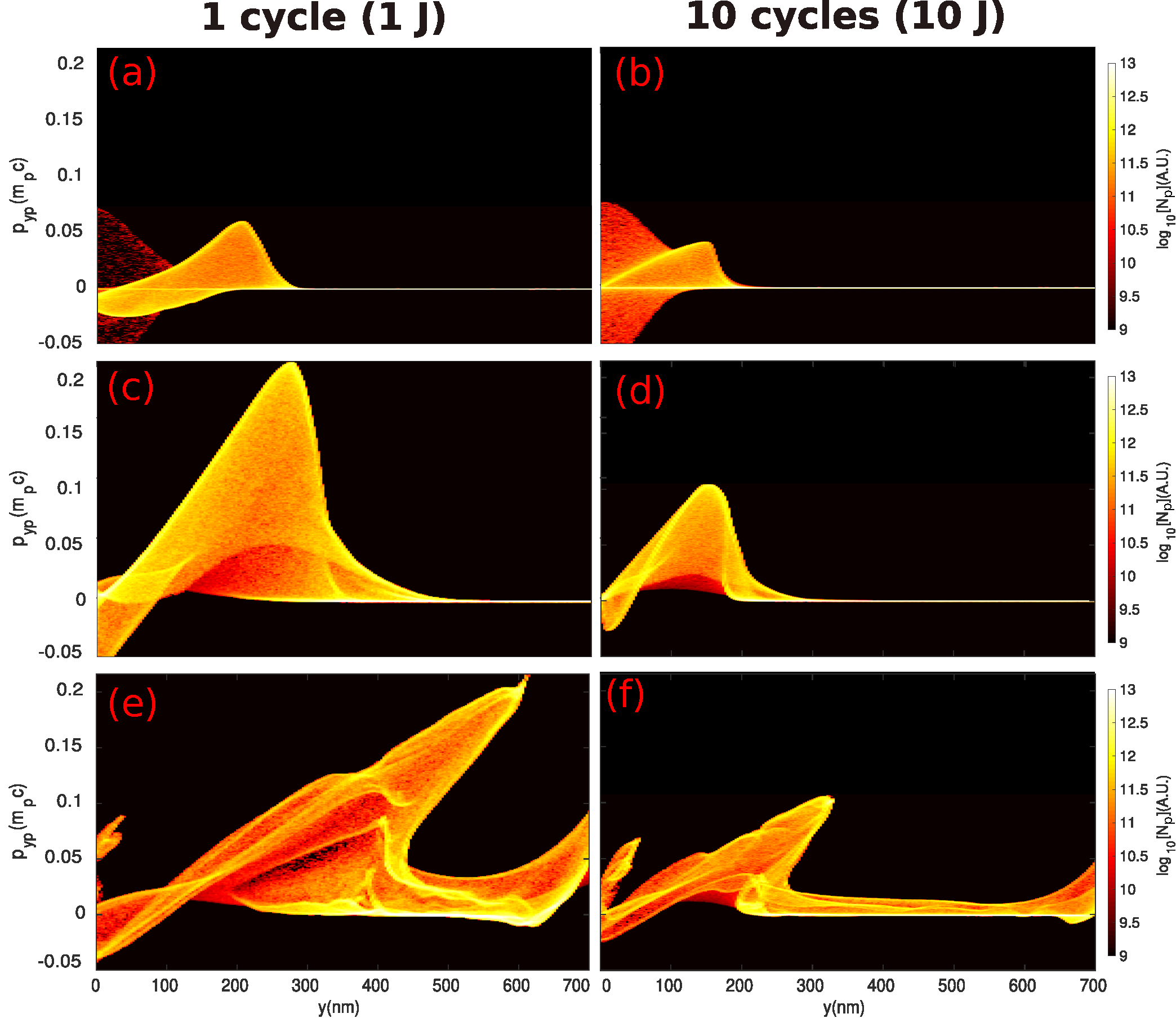}
    \caption{Proton phase spaces at (a)--(b) $t=50~T_L$, (c)--(d) $t=90~T_L$ and (e)--(f) $t= 250~T_L$. (a)--(c)--(e) with a single--cycle laser pulse. (b)--(d)--(f) with a 10--cycles laser pulse.}\label{proton_phase_space}
\end{figure}

When the accelerated ion density peak reaches the outward edge of the electron compression layer ($y_i \approx l$), the charge separation field of the electron compression layer drops to zero (see Eq. \eqref{ion_eq}). As consequence, the ion density spike field is no longer dynamically screened by the electrons which causes hydrodynamical breaking. The ion density spike splits up rapidly (see Fig. \ref{1D_cuts}(b)) and a short bunch of fast ions is generated \cite{Macchi_2009}. The onset of hydrodynamical breaking causes the formation of an ambipolar sheath field with a sharp gradient, around the breaking point (i.e. $y_i \approx l$) as well as a strong electron heating (see Fig. \ref{radial_energy_spectra}(c)) which induces electron density modulations (see Fig. \ref{1D_cuts}). The slower ions are reflected by the negative part of the ampibolar sheath field and acquire negative radial velocity in the frame of reference of the ion density spike. The related amount of energy can be estimated as $\sim e E_s d$ where $E_i$ is the maximal amplitude of the sheath field and $d$ is the width of the ion peak. From simulation results we have $d \simeq0.3~ \lambda_L $ and $E_i \simeq 0.003~E_L$, we obtain a characteristic energy $eE_id \simeq0.5$ MeV. Given the energies involved, the energy given by this field as a boost is negligible for energetic ions. However, the negative part of the ampibolar electric field tends to slow down the ions which have kinetic energies less than the sheath field characteristic energy and, and thus separates the slowest ions from  the fastest. This can be identified in Fig. \ref{proton_phase_space}(c) via a breaking in the shape of the front of the ion phase space at $y \simeq 350$ nm and for $p_{yp}\simeq 0.04~m_ic$. 


\section{A path to proton--Boron fusion using high energy attosecond pulses}\label{Sec4}
Using these interesting and encouraging simulation results, we propose a path to proton--Boron fusion using ultra-intense attosecond pulses. We indeed observe in our numerical simulations that in a first step, protons are accelerated radially in the MeV range by the propagation of the "ultra-intense" attosecond pulse in the solid target. In a second step, these protons propagate in the surrounding target and interact with immobile Boron ions to produce alphas (3 alphas for each pB fusion reaction). In a third step, the produced alphas can interact with the surrounding protons and Boron ions to accelerate them and launch new fusion reactions.

To evaluate the alpha production from proton--Boron fusion, we can use the same estimate that has been proposed by Giuffrida~\cite{Giuffrida}. This model takes into account the total proton--Boron fusion cross section (see Fig.~\ref{cross_section}) and the stopping power of ions in solid targets. The number of $\alpha$ particles produced ($N_\alpha$) is linked to the proton number ($N_p$) and the Boron number ($N_B$) entering in the target by$:$
\begin{equation}
    N_\alpha = 3~N_p~P_p+3~N_B~P_B,\label{Na}
\end{equation}
where $P$ is the reaction probability and can be written$:$
\begin{eqnarray}
  P_p = n_{B0} I_p(E_{p0}),\\
  P_B = n_{p0} I_B(E_{B0}), 
\end{eqnarray}
where $n_{B0}$ and $n_{p0}$ are the initial proton and Boron number density inside the target and $I_p$ and $I_B$ is defined by$:$
\begin{eqnarray}
  I_p(E_{p0})=\int_0^{E_{p0}}{\sigma_{pB}(E)\left(\frac{dE}{dx}\right)_p^{-1}~dE},\\
  I_B(E_{B0})=\int_0^{E_{B0}}{\sigma_{pB}(E)\left(\frac{dE}{dx}\right)_B^{-1}~dE}, \label{Ip}
\end{eqnarray}
where $E_{p0}$ and $E_{B0}$ are the initial proton and Boron energies, respectively and $\sigma_{pB}$ and $dE/dx$ are the total proton--Boron fusion cross section and the stopping power of the corresponding ions, respectively. The stopping powers are calculated with the Monte--Carlo code SRIM~\cite{SRIM} and the pB fusion cross section is given in Ref.~\cite{Nevins_Swain}.
\begin{figure}
    \includegraphics[scale=0.33,angle=0,clip]{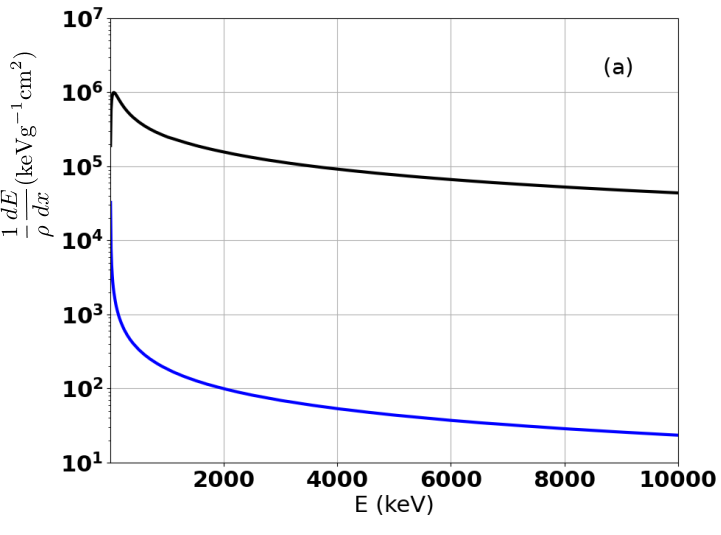}
    \includegraphics[scale=0.33,angle=0,clip]{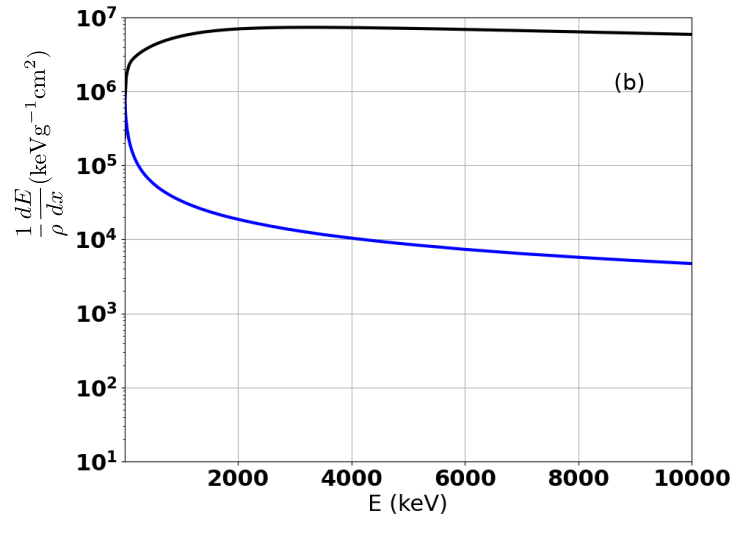}
    \caption{Mass stopping power as function of the energy in a pB (proton--Boron) solid target. In black Electronic mass stopping power and in blue the nuclear mass stopping power. (a) in the case of transport of proton in pB. (b) In the case of transport of Boron in pB.}
    \label{Stopping}
\end{figure}

Now let's consider ions accelerated by attosecond pulses described in the previous numerical PIC simulations (see section \ref{Sec3}).

In the case of a 10--cycle pulse$:$ proton and Boron ions are accelerated radially at $\sim$4 MeV and $\sim$10 MeV respectively (see Fig.~\ref{radial_plots}). Considering energy efficiency production for proton 0.0057 $\%$ and for Boron 0.013$\%$ (see Table \ref{ion_absorptions}, from equations~\eqref{Na}--\eqref{Ip}, we can estimate the ratio between the energy produced by fusion reactions and the attosecond pulse energy. In this case, we obtain a ratio of 3$\times10^{-8}$.

\begin{center}
\begin{table}[h]
\caption{Laser energy converted into radial ion kinetic energy.}
\setlength{\arrayrulewidth}{0.3mm}
\setlength{\tabcolsep}{7pt}
\renewcommand{\arraystretch}{2}
\begin{tabular}{|c|c|c|}
   \Xhline{2\arrayrulewidth}
  &  \multicolumn{1}{c|}{single--cycle} & 10--cycles \\
  \hline
  Laser energy & 1~J & 10~J\\
    \hline 
    Proton & 0.085~\% & 0.0057~\% \\
    Boron  & 0.2~\% & 0.013~\% \\
    \Xhline{2\arrayrulewidth}
\end{tabular}\label{ion_absorptions}
\end{table} 
\end{center}

In the case of the 1~J single--cycle pulse$:$ proton and Boron ions are accelerated transversely at $\sim$ 20~MeV and $\sim$~45 MeV, respectively. Considering from PIC simulations, the proton and Boron energy efficiencies are 0.085~$\%$ and 0.2~$\%$, respectively. The ratio between fusion energy and laser energy becomes 8$\times10^{-7}$. This is 25 times higher than with a 10--cycle pulse (10~J), with ten times less energy. These results concerning the single--cycle pulse are encouraging to pursue this study to increase the alpha production efficiency as the energy efficiencies are only estimated on the duration of the numerical simulation i.e. about 10~fs. In all these previous results we have taken into account only the stopping power in ordinary matter, based on the Bethe--Bloch formula with corrections and only solid matter stopping power has been taken into account (see the black curve in Fig.~\ref{Stopping}). In the case of a plasma state for the target, the stopping power is different and decreases (see Ref.~\cite{Giuffrida}) and is situated between the two curves in Fig.~\ref{Stopping} where the blue curve gives the nuclear stopping power ("totally ionized matter", i.e. without electron) and the black curve the Bethe--Bloch stopping power for solid target. In this case of totally ionized plasma, the stopping power is 1000 times less and it follows that the alpha production becomes 1000 times more important. This situation may be possible if fusion reactions take place inside the channel where the electron density is negligible and could increase the alpha production. Therefore, if fusion reactions take place inside the channel, the alpha production will be increased due to the domination by the nuclear stopping power. 

Other effects can contribute to increase the production of alpha particles. For an attosecond single--cycle pulse with very high intensity, and a very high electric field, the tunneling effect in the fusion reaction can be enhanced. In a very recent work~\cite{Kohlfurst}, time-dependent tunneling rates were studied for the proton--Boron fusion reaction. They found that for a pulse in the keV energy range and a field peak of~$10^{16}$~V/m, the tunneling rate is significantly enhanced (one order of magnitude) in the 40--80~keV energy range. These electric field magnitudes are quite close to the attosecond pulse considered in the study presented here. Moreover, in a recent work, Shmatov~\cite{Shmatov} drew attention to recent pB cross-section measurements at the Triangle Universities Nuclear Laboratory \cite{Sikora}. They have shown that the cross-section at the 600~keV resonance is 17~$\%$ higher and up to 7~times higher in a several MeV range than the Nevins\&Swain reference data \cite{Nevins_Swain}. 

Taking into account the increase in the number of fusion reactions in the channel, the increase of the cross-section mentioned above, and using a more realistic propagation length (1D PIC simulations not shown here show a depletion length of $\sim$ 16 microns in this case i.e. 32 times longer than the target thickness in our 2D PIC simulation), the ratio between the energy produced by fusion reactions and the attosecond pulse energy becomes 0.03 for the single--cycle case, which is 375 times higher than the state--of--the--art results obtained by Giuffrida \textit{et al.}~\cite{Giuffrida}. If these results are confirmed, this will be very encouraging for the feasibility of pB fusion. 

\section{Discussion and conclusion}\label{Sec5}
Ultra--intense attosecond pulses have the potential to revolutionize ultra--high intensity laser--plasma interaction. Tremendous progress is expected for many applications such as compact particle accelerators, extreme laboratory astrophysics and strong field quantum electrodynamics. In this article, we have proposed the use of these attosecond pulses to increase the efficiency of proton--Boron fusion. First, we have investigated the interaction of the pulse with a solid density proton--Boron target using 2D PIC simulations. We have demonstrated that using a single--cycle pulse, protons and Boron ions can be efficiently accelerated radially to energies that correspond to the maximum of the proton--Boron fusion cross section. Our findings are supported by analytical estimates explaining the advantage of single--cycle high energy attosecond pulses. For durations much longer than the laser period, the electrons are expelled by the gradients of the laser pulse and a mechanical equilibrium between the transverse ponderomotive force and the radial electrostatic force can be established. This determines the characteristic radius of the channel as well as the (maximum) energy of radially accelerated ions. When a single--cycle laser pulse is used, the situation is drastically different. The radial ejection of electrons from the highest field regions is now primarily due to a \textit{kick} induced by the laser electric field itself. The radius of the channel (which is approximately the electron radius excursion) turns out to be larger (about 50\% more) compared with the case of a 10--cycle laser pulse. Consequently, the ions are radially accelerated over a longer distance and thus gain more kinetic energy. Especially, the energy for the fastest ions is multiplied by 4 compared with the case using a 10--cycle laser pulse.  

We have then estimated the number of produced alphas by proton--Boron fusion reactions using the characteristics of the protons and of the Boron ions obtained in our PIC simulations. Our results are very encouraging and demonstrate that attosecond proton--Boron fusion has already the potential to obtain similar results as more complicated and costly fusion paths such as Magnetic Confinement Fusion and Inertial Confinement fusion \cite{Patent}.

In order to progress towards attosecond proton--Boron fusion the influence of self--generated magnetic fields on the number of produced alphas, the influence of the shock structure on longer time scales, and its possible transition to a detonation should all be investigated. The problem is quite complex because the initial physical processes are at a kinetic spatio--temporal scale and the final stage is at a hydrodynamic spatio--temporal scale. The problem is very challenging but very exciting because it has not been explored up--to now and could lead to promising side applications. 

These questions could be treated numerically using PIC simulations including proton--Boron fusion reactions and collisions to obtain the number of alpha particles that can be produced and the heating of the pB solid target around the interaction volume. The next step corresponding to the heating of the target on longer time scales and the production of new fusion reactions using MHD simulations will use the results of the PIC simulations as inputs. The number of produced alphas could also be improved by structuring the target or using innovative materials.
\begin{acknowledgments}
This research was supported by the French National Research Agency (Grant No. ANR-17-CE30-0033-01) TULIMA Project. We thank P. Chen, H. Vincenti, N.~Naumova for the interest they’ve shown in this study and their advice.

\end{acknowledgments}


\begin{thebibliography}{00}
\bibitem{comment} Strictly, we cannot speak of \textit{ponderomotive force} when the number of cycles of the electromagnetic pulse is not large as the oscillation center of motion does not exist \cite{Bauer}.
\end{thebibliography}
\end{document}